\documentclass[aps,twocolumn,superscriptaddress,longbibliography,pre]{revtex4-1}
\usepackage{amssymb}
\usepackage{bm,amsmath}
\usepackage{graphicx} 
\usepackage{array}
\usepackage{setspace}
\usepackage{xcolor}
\usepackage{hyperref}
\usepackage{gensymb}
\newcommand{\blue}{\textcolor{blue}}
\hypersetup{
	colorlinks,
	linkcolor={red!90!black},
	citecolor={black!10!blue},
	urlcolor={blue!80!black}
}

\renewcommand{\r}{\textcolor{black}}

\begin{document}
\title{Fluctuation-dominated phase ordering in the one dimensional Truncated Inverse Distance Square Ising (TIDSI) model}
	\author{Souvik Sadhukhan}
	\email{ssadhukhan@tifrh.res.in}
	\affiliation{TIFR Centre for Interdisciplinary Sciences, Tata Institute of Fundamental Research, Hyderabad - 500046, India}
	\author{Mustansir Barma}
	\email{barma@tifrh.res.in}
	\affiliation{TIFR Centre for Interdisciplinary Sciences, Tata Institute of Fundamental Research, Hyderabad - 500046, India}
    \author{Saroj Kumar Nandi}
    \email{saroj@tifrh.res.in}
    \affiliation{TIFR Centre for Interdisciplinary Sciences, Tata Institute of Fundamental Research, Hyderabad - 500046, India}

\begin{abstract}

Many physical systems, including some examples of active matter, granular assemblies, and biological systems, show fluctuation-dominated phase ordering (FDPO), where macroscopic fluctuations coexist with long-range order. Most of these systems are out of equilibrium. By contrast, a recent work has analytically demonstrated that an equilibrium one-dimensional Truncated Inverse Distance Square Ising (TIDSI) model shows FDPO. The analytical results rely on a cluster representation of the model that we term TIDSI-CL and are governed by the ratio, $c$, of the long-range interaction strength to the critical temperature. We show that the allowed range of $c$ is very narrow in the original TIDSI model while it is unbounded in TIDSI-CL. We perform Monte-Carlo simulations for the TIDSI model and show consistency with the analytical results in the allowed range of $c$. The correlation length grows strongly on approaching the critical point, leading to a broad near-critical region. Within this region, $\alpha$, which is the cusp exponent of the power-law decay of the scaled correlation function at criticality, changes to $\alpha^\text{eff}$. We also investigate the coarsening dynamics of the model: the correlation function, domain size distribution, and aging behavior are consistent with the equilibrium properties upon replacing the system size, $L$, by the coarsening length, $\mathcal{L}(t)$. The mean largest cluster size shows logarithmic corrections due to finite $L$ and waiting time, $t_w$. The aging autocorrelation function exhibits two different scaling forms, characterized by exponents $\beta$ and $\gamma$, at short and long times compared to $t_w$, where $\beta=\alpha/2$.

\end{abstract}
\maketitle

\section{Introduction}
Over the past two decades, fluctuation-dominated ordered states, where macroscopic fluctuations coexist with long-range order, have been observed in several models of non-equilibrium systems. These include particles on a fluctuating surface \cite{Das2000, Das2001, Manoj2003, Nagar2005, Chatterjee2006, Kapri2016,Mahapatra2020,Das2023,barma2023book}, active particles \cite{Mishra2006,Dey2012}, vibrating rods  \cite{Narayan2007}, freely cooling granular colliding systems \cite{Shinde2007}, and actin clustering on the cell surface \cite{Das2016}. These systems show a broad distribution of the order parameter without losing macroscopic order in their steady states \cite{Das2001, Chatterjee2006, Das2016}; this indicates giant fluctuations, even in the thermodynamic limit. In addition, the scaled two-point spin-spin or density-density spatial correlation function displays a distinctive cusp singularity at small values of the scaled separation. These are the two chief characteristics of states which show fluctuation-dominated phase ordering (FDPO).

Usually, in the steady state of systems with a conserved order parameter, the ordered state is phase separated; each phase occupies a macroscopic region in the system \r{and phases are separated by sharp interfaces. While coarsening, the interface is sharp compared to the coarsening length $\mathcal{L}(t)$, and consequently, the spatial correlation function, $G(r,t)$, decays linearly with the separation $r$: $G(r,t)\sim 1-2r/\mathcal{L}(t)$. This leads to the Porod's law \cite{Porod1952,Bray1994} according to which the structure factor $\sim (k\mathcal{L})^{-(d+1)}$, where $d$ is the spatial dimension and $k$ is the wave vector. Once the steady state is reached, we may replace $\mathcal{L}$ by the system size $L$ \cite{Das2001} and $G(r,L)$ decays as $1-2r/L$.}
	
\r{By contrast, in systems exhibiting FDPO, the interface is broad and $G(r,L)$ takes the steady state form}
\begin{equation}
	C(r)\sim 1-A\left|\frac{r}{L}\right|^\alpha,
	\end{equation}
where $\alpha$ $(0<\alpha<1)$ is the cusp exponent. The steady state of such systems shows large domains of the ordered phases separated by large interfacial disordered regions. The largest of the ordered domains is of the order of the system size, consistent with long-range order. Moreover, the disordered regions also scale with the system size leading to the violation of the Porod's law. Most of the systems exhibiting FDPO have nonequilibrium steady states; it is only recently that the presence of FDPO has been demonstrated \cite{Barma2019} in an equilibrium system \cite{Bar2014, Bar2014PRL, Bar2016}.

Studies of phase transitions generally distinguish first order discontinuous transitions from the second order continuous ones \cite{kardarbook}. \r{However, some systems show mixed order transitions (MOT) with some characteristics of both. Early examples of MOTs include} the long range inverse distance square Ising (IDSI) model \cite{Anderson1969,Thouless1969,dyson1971,corberi2021} and models of depinning transitions such as DNA denaturation \cite{fisher1966,Douglas1966}. To establish a link between these two classes of models and study the MOT in the resulting model, Bar and Mukamel introduced a truncated inverse distance square Ising (TIDSI) model in which long-range interactions act only within domains (or clusters) of like spins \cite{Bar2014, Bar2014PRL,mukamel2023book}. This model shows MOT along a critical locus, from a high temperature disordered phase to a low-temperature ordered phase: on the one hand it exhibits an algebraically diverging correlation length in the disordered phase, as in a second-order transition, and on the other hand shows a discontinuity in the order parameter as in a first-order transition \cite{Bar2014, Bar2014PRL,mukamel2023book}. 
\begin{figure}[h!] 
	\centering
	\includegraphics[width=7.2cm]{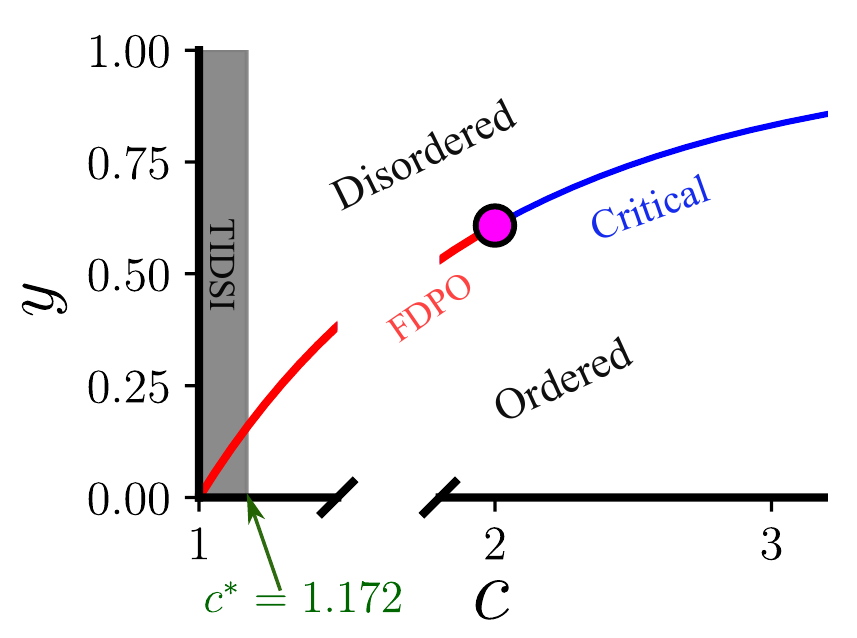}
	\caption{Phase diagram of the TIDSI-CL model in the ($c-y$) plane, while $c > 1$ and $y$ is the fugacity of the domain walls (defined in Sec.~\ref{critline}). The disordered paramagnetic phase ($y > y_c$) and ordered ferromagnetic phase ($y < y_c$) are separated by the critical line, $y_c = 1/\zeta(c)$. At $c = 2$ (magenta dot), the behaviour of TIDSI-CL changes from normal critical (blue line) to FDPO (red line). The TIDSI model is defined within the gray shaded region up to $c^*$. The broken line denotes the continuation of the curve.}
	\label{phasespace}
\end{figure}

A consequence of the truncated interactions is that instead of summing over spin configurations, the partition function can alternatively be rewritten as a sum over cluster configurations $\{l_n\}$ \cite{Bar2014, Bar2014PRL, Bar2016, Barma2019}. We call this representation the TIDSI-Cluster or TIDSI-CL model. The distinction between TIDSI and TIDSI-CL is important; as shown below, the latter is defined in a broader range of parameters than the former. In the theory, a key role is played by $c$, the ratio of the strength of the long-range interaction $C$ to the critical temperature $T_c$.  In the TIDSI model, since $T_c$ is proportional to $C$ for very large $C$, the ratio $c = C/T_c$ is bounded above by a certain value, say $c^{*}$ [see Fig.~(\ref{phasespace})]. By contrast, in the TIDSI-CL model, $c$ appears as a parameter weighting cluster lengths, and its range is unrestricted. The advantage of the cluster representation is that many quantities of interest can be calculated analytically. For instance, for $1 < c < 2$, the length of the largest cluster was shown to be extensive and to have a distribution that remains broad even in the thermodynamic limit, while for $c > 2$, it is sub-extensive \cite{Bar2016}. Further, along the critical line for $1 < c < 2$, the two-point correlation function is a function of $r/L$, where $r$ is the separation and $L$ is the system size. It shows a cusp singularity, $1-A|r/L|^\alpha$, at small scaled distances with a cusp exponent $\alpha$ ($0< \alpha < 1$). On the other hand, it shows a simple power-law decay $1/r^{d-2+\eta}$ for $c > 2$ \cite{Barma2019,barma2023book,mukamel2023book}. 

In this paper, we work with the TIDSI model in the spin representation and perform Metropolis Monte Carlo (MC) simulations using single-spin-flips. Besides validating the analytic results obtained within the TIDSI-CL framework for static, equilibrium properties within the physical range for the TIDSI model $(c < c^*)$, we also obtain new results for time-dependent spin correlation functions, both in equilibrium and during coarsening. 

Our primary findings can be summarized as follows:

(1) The allowed parameter range of the TIDSI model is $1 < c < c^*$ with $c^* \simeq 1.172$. Within this accessible range, the TIDSI-CL model gives a good description of the properties of the TIDSI model.

(2) \r{Our simulation data for the distribution of the size of the largest domain $l_\text{max}$ in steady state at $T_c$ agrees well with the analytic results \cite{Bar2016}. We show that $l_\text{max}$ exhibits} large fluctuations typical of FDPO along the critical locus in the accessible range $1 < c < c^*$.

(3) During coarsening following a quench to the critical point, we find that at time $t$, correlations develop within a length scale $\mathcal{L}(t)$ which grows as $\mathcal{L}(t) \sim t^{1/z}$ with $z = 2$. The globally largest domain is proportional to $\mathcal{L}(t)$ with a multiplicative logarithmic correction involving $L/\mathcal{L}(t)$, as shown recently in \cite{Das2023}. Further, the two-point correlation function during coarsening exhibits scaling as a function of $r/\mathcal{L}(t)$, and displays a cusp with the same exponent as predicted analytically \cite{Barma2019, barma2023book}.

(4) In the disordered phase, the correlation length $\xi$ is very large in the range $[1,c^*]$ if $T\gtrsim T_c$. Thus, there is a very large finite size effect if, as happens in practice, the condition $\xi >> L$ holds. Then, if the system is quenched to temperatures slightly above $T_c$, the correlation function $G(r)$ continues to show a cusp as a function of $r/\mathcal{L}(t)$. However, the cusp singularity is now described by an effective exponent,  $\alpha^{\text{eff}}$, which is distinct from the analytic value $\alpha$ predicted by TIDSI-CL. On the other hand, if the system is quenched below $T_c$, we observed that $\alpha^{\text{eff}}$ approaches $1$ as expected for normal phase ordering.

(5) We studied the time auto-correlation function of the TIDSI model and found that in steady state it is a singular scaling function of $t/L^z$, characterized by a cusp exponent $\beta = \alpha/z$ with $z = 2$. 
Next, we studied the steady-state properties of the auto-correlation function of the TIDSI model for a fixed $T$ ($\geq T_c$). We extracted the cusp exponent from the scaled correlation function at steady state and found that it agrees with the corresponding $\alpha^{\text{eff}}$ during coarsening at the same $T$.

(6) We also computed the aging auto-correlation function, $G(t,t_w)$ which is a scaling function of $t_w$ and $t$ respectively, where $t_w$ is the waiting time and $t$ is the time difference. For $t >> t_w$, $G(t,t_w)$ decays algebraically with a power $\gamma$ that increases monotonically with $c$. Moreover, for $t << t_w$, $G(t,t_w)$ shows a cusp singularity with an exponent $\beta$ as a function of $t/t_w$, similar to the autocorrelation function in steady state. We verified that $\beta = \alpha/2$ during coarsening as well as in steady state.

\section{Model}\label{modeldes}
The TIDSI (truncated inverse distance square Ising) model is a variant of the well-known inverse distance square Ising (IDSI) model in one-dimension. The energy function governing the behavior of the IDSI model \cite{Thouless1969} is
\begin{equation}
	\mathcal{H}=-\sum_{i<j}J(j-i)S_i S_j; \,\,\, J(r>>1)\sim 1/r^2,
	\end{equation}
	where $i$ and $j$ corresponds to spatial positions on a lattice of length $L$, $S_i=\pm 1$, and $J(r)\geq0$ for all $r$.
Note that the exponent 2 of the long-range interaction is crucial in spatial dimension one. \r{The IDSI is a special case of the $1D$ Ising model with power-law decaying interactions, $J(r)\sim 1/r^{1+\sigma}$, and separates the region with no order $(\sigma>1)$ from that with long range order below a critical temperature ($\sigma<1$) \cite{dyson1971,frohlich1982,tomita2008}. The IDSI model itself
 shows a Kosterlitz-Thouless type transition \cite{Bar2014,frohlich1982}.}

\begin{figure}
	\centering
	\includegraphics[width=8.6cm]{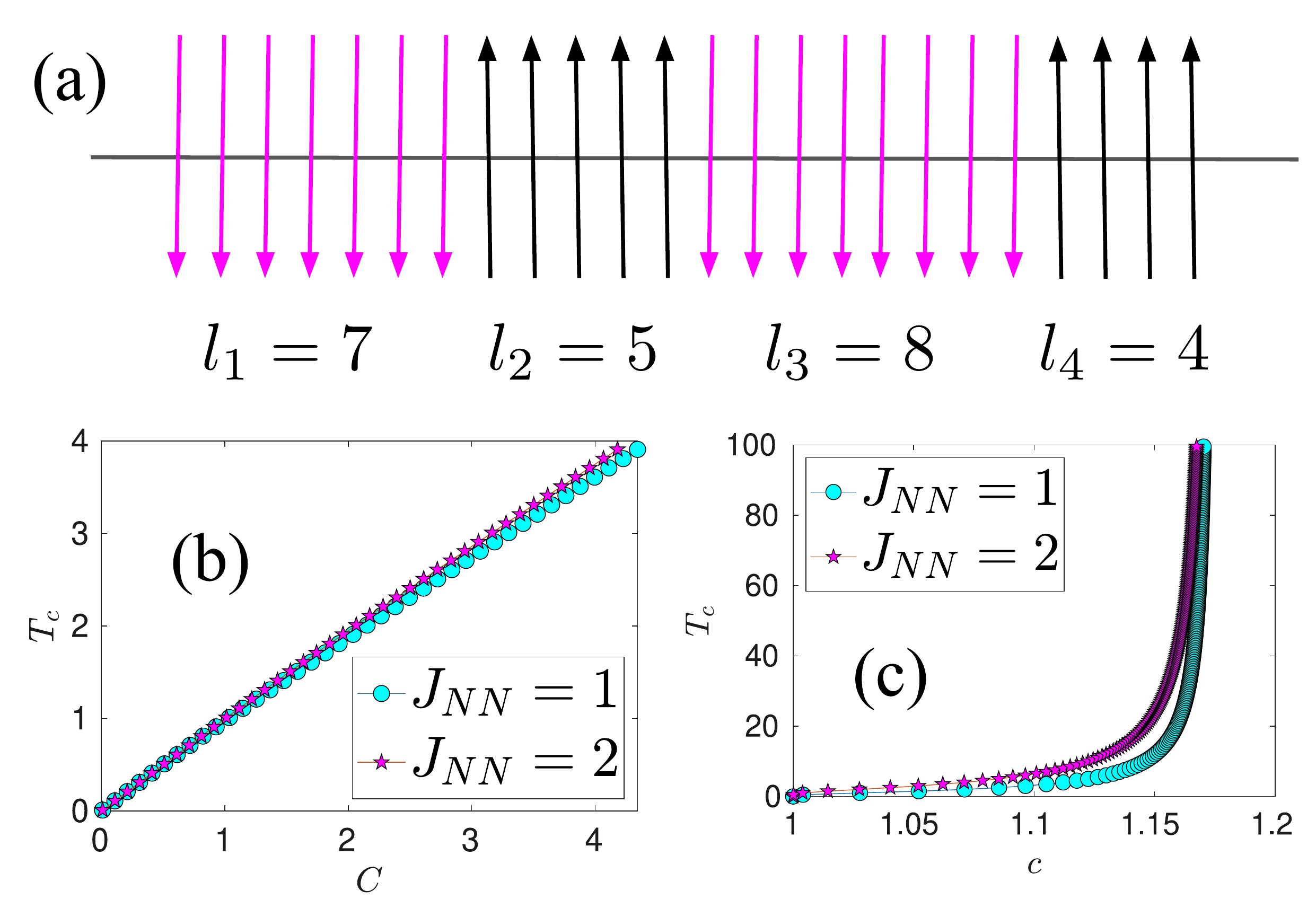}
	\caption{TIDSI model and its critical line. (a) A schematic representation of the TIDSI model with four typical domains. The long-range interaction is limited to each domain.  (b) The critical line in the $T_c - C$ plane. $T_c$ becomes linearly proportional to $C$ at large $C$; this leads to the upper bound for $c = C/T_c$. (c) Plot of the critical line in $T_c -c$ plane. $T_c$ diverges when $c$ approaches $1.172$ in the TIDSI model.}
	\label{model}
\end{figure}

When we truncate the long-range interaction of the IDSI model by limiting it to act within domains, we obtain the TIDSI model \cite{Bar2014,Bar2014PRL}.
A typical configuration of the model consists of a succession of domains, defined as stretches of like spins (Fig. \ref{model}a). Thus, the Hamiltonian describing the model in one dimension is
\begin{equation}\label{Hamiltonian}
\mathcal{H} = -\sum_{\langle i,j \rangle }J_{NN} S_i S_j - \sum_{i<j} J(i-j) S_i S_j I(i\sim j)
\end{equation}
where $\langle i,j \rangle$ denotes nearest neighbor pairs, and $i$ and $j$ run from 1 to $L$, the total number of spins. $J_{NN} > 0$ is the ferromagnetic nearest neighbour interaction, while $J(i-j) = \frac{C}{(i-j)^2}$ varies with distance and  
\begin{equation}
I(i\sim j) =
\begin{cases}
1, &         \text{if $i,j \in $  same domain}\\
0, &         \text{otherwise}.
\end{cases}
\end{equation}
$I$ is a cut-off function which ensures that the long-range interaction acts only when $i$ and $j$ belong to the same domain. Explicitly, in terms of the spin variables, we have
	\begin{equation}
		I(i \sim j) = \prod_{k=i}^{j-1} \frac{1 + S_k S_{k+1}}{2}.
	\end{equation}
The rescaled parameter $c=C/T_c$, where $T_c$ is the critical temperature, governs the properties of the TIDSI model \cite{Bar2014PRL,Barma2019}.

As shown in Refs. \cite{Bar2014PRL,Bar2014,Bar2016} and in Eq. (\ref{effectiveHamiltonian}) below, we can express Eq. (\ref{Hamiltonian}) in terms of the cluster lengths, $\{l_n\}$. The analytic treatment of Ref. \cite{Barma2019} was carried out using this cluster representation and the results are presented in terms of $c\geq1$. Within this representation, there is no upper bound on $c$ as it can be treated as an independent parameter. However, using the microscopic parameters with $J_{NN}>0$ leads to an upper bound $c*$ on the value of $c$ (Sec. \ref{critline}); higher values of $c$ would require $J_{NN}<0$. We call the model TIDSI-CL when we refer to the analytical results with the cluster representation and unbounded $c$ whereas we reserve the term TIDSI for the microscopic model, Eq. (\ref{Hamiltonian}), with $J_{NN}>0$, and consequently, with $c<c^*$. 

To the best of our knowledge, so far TIDSI is the only equilibrium model known to exhibit FDPO \cite{Barma2019}. The coarse-grained depth (CD) model, 
which is connected to a problem of sliding particles on an infinitely slowly evolving fluctuating surface,
and which shows FDPO, 
has been shown to be a special case of the TIDSI-CL model \cite{Das2023}. We emphasize the distinctive natures of the TIDSI model compared to the IDSI model: The TIDSI model shows a power-law divergence of the correlation length $\xi$ in the disordered phase whereas the IDSI model shows a much faster divergence, $\xi\sim \exp[\sqrt{1/(T-T_c)}]$ \cite{Bar2014,Bar2014PRL,barma2023book,mukamel2023book}. Moreover, the critical locus in the TIDSI model has a part on which the system exhibits FDPO \cite{Barma2019}.

We have simulated the TIDSI model using the single spin-flip Metropolis Monte Carlo (MC) algorithm: the probability of flipping a spin is $P = \text{min}(1, e^{-\Delta\mathcal{H}/T})$, where $\Delta \mathcal{H}$ is the energy change due to the spin-flip. We have used the cluster finding algorithm, generally used in the Wolff cluster algorithm, to identify and distinguish domains during the simulation \cite{wolff1989}. We have studied both free and periodic boundary conditions. There is a slight difference between the two as explained below.

Let us consider a spin configuration with $N$ domains, $1 \le N \leq L$, and domain lengths $\{l_i\}_{i=1}^N$, where $l_i \ge 1$. Furthermore, $\sum_{i=1}^Nl_i = L$. In the case of free boundary conditions, there is no restriction on $N$, and we can write $\mathcal{H}$ in terms of the domain lengths as
\begin{equation}\label{FBCHamiltonian}
\mathcal{H} = -J_{NN}-J_{NN}\sum_{n=1}^{N}(l_n - 2)  - \sum_{n=1}^{N} \sum_{r = 1}^{l_n}(l_n - r)J(r).
\end{equation}

By contrast, for periodic boundary conditions (PBC), $N$ can only be even, except for $N=1$, the fully ordered state. In that case, the short-range part of $\mathcal{H}$ changes,
\begin{equation}\label{pbcdomainH}
\mathcal{H} = \sum_{n=1}^{N}-J_{NN}(l_n - 2) - \sum_{n=1}^{N} \sum_{r = 1}^{l_n}(l_n - r)J(r).
\end{equation}
Note that the short-range energy differs for the specific case of $N=1$, where it is $-LJ_{NN}$. This slight difference in $\mathcal{H}$ for the two boundary conditions affects the transition temperature for a finite system. However, from our simulations for both types of boundary conditions, we find that, except for the value of $T_c$, the qualitative results do not change (see Appendix \ref{comp} for an illustration). Therefore, we will present our simulation results only for periodic boundary conditions.

\section{Results}

We present our simulation results for the TIDSI model in one dimension: first for the coarsening dynamics and then for the dynamics in equilibrium. Note that by dynamics, we mean the time-dependent properties that we study via the MC simulation. However, before that, it is instructive to compare the critical line of the TIDSI model with that of the TIDSI-CL model.

\subsection{The critical line for the TIDSI and TIDSI-CL model}
\label{critline}

The TIDSI Hamiltonian can be written in terms of domain lengths $\{l_i\}$ as follows. Following Refs. \cite{Bar2014} and \cite{Bar2016}, we replace the sums in Eq. (\ref{Hamiltonian}) by integrals as would be valid for large $l_i$. We find
$\sum_{r = 1}^{l_n} J(r) \approx a - \frac{C}{l_n} + O(l_n^{-2})$ and $\sum_{r = 1}^{l_n} rJ(r) \approx b + {C} \ln({l_n}) + O(l_n^{-1})$, where $a$ is a constant and $b$ is Euler's constant times $C$. For large domains, we may neglect the $O(l_n^{-1})$ term and rewrite Eq. (\ref{FBCHamiltonian}) as
\begin{equation}\label{effectiveHamiltonian}
\mathcal{H}_{CL} = C \sum_n \ln(l_n) + N\Delta,
\end{equation}
where $\Delta = 2J_{NN} + C + b$ and $ c = \beta C$.
The subscript $CL$ is used as $\mathcal{H}_{CL}$ involves only the cluster lengths $\{l_i\}$. On defining the fugacity $y = \exp(-\beta \Delta)$, the expression for the corresponding partition function \cite{Barma2019} is 
\begin{equation}
	\mathcal{Z}_{CL}=\sum_{n=1}^\infty y^N \sum_{l_1=1}^\infty \ldots \sum_{l_N=1}^\infty \prod_{n=1}^N \frac{1}{l_n^c}\delta_{\sum_{n=1}^Nl_n,L}.
\end{equation} 
Analysis of $\mathcal{Z}_{CL}$ leads to the conclusion that there is a critical locus in the $y-c$ plane given by the critical line as $y_c=1/\zeta(c)$, where $\zeta(c)$ is the Riemann zeta function, and $c=C/T$, where we have set the Boltzmann constant to unity \cite{Barma2019}.

We plot the critical temperature $T_c$ as a function of $C$ in Fig. \ref{model}(b) and as a function of $c$ in Fig. \ref{model}(c) for two values of $J_{NN}$. Figure \ref{model}(b) shows that $T_c$ has a nearly linear behavior at large $C$ in the $T_c-C$ plane implying that $\beta_c C$ approaches a constant. Consequently, $T_c$ diverges at a critical value of $c\simeq 1.172$ (Fig. \ref{model}c). Thus, in the spin representation of the TIDSI model with ferromagnetic interactions, $c$ cannot exceed $1.172$, while it can take on any value in the cluster representation of the TIDSI model, TIDSI-CL. Below we test the analytic predictions against simulations of the TIDSI model for $c\leq 1.172$.

\begin{figure*}
	\centering
	\includegraphics[width=18cm]{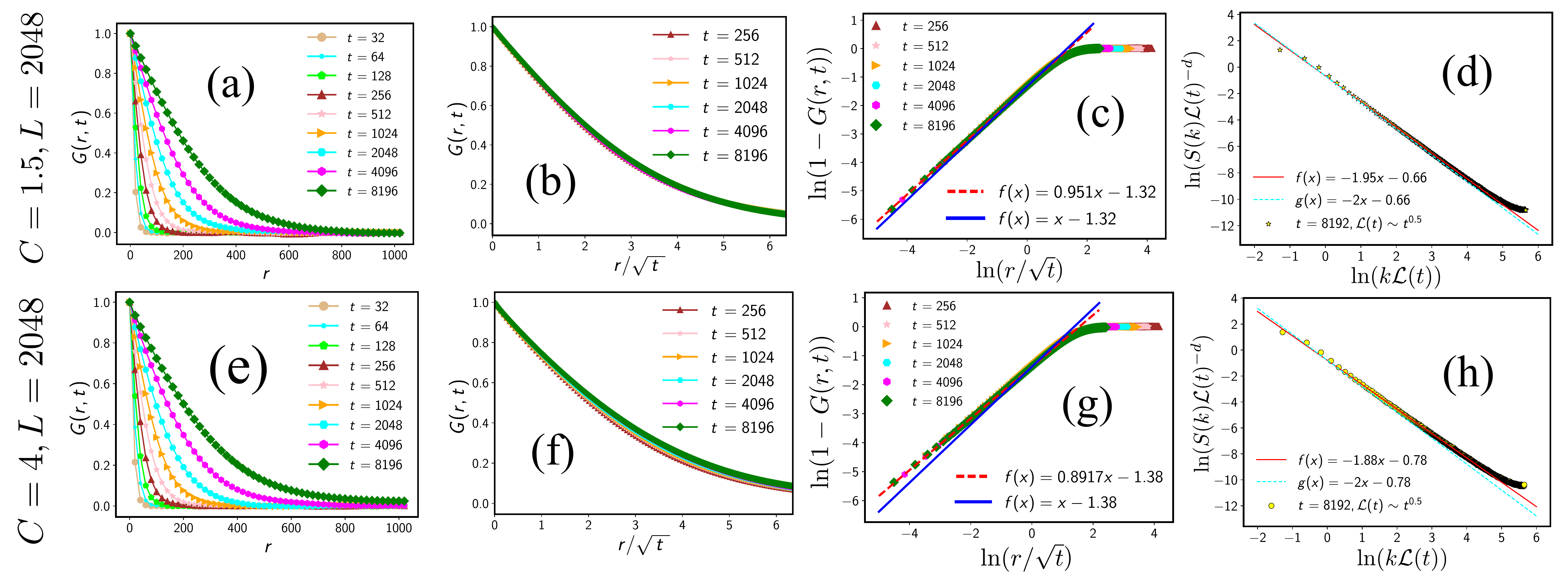}
	\caption{Spin-spin spatial correlation functions in the coarsening regime. (a) $G(r,t)$ as a function of $r$ for different values of $t$. (b) Scaling collapse of $G(r,t)$ when plotted as a function of $r/\sqrt{t}$. (c) Fitting the early part of the data of $\ln(1-G(r,t))$ as a function of $\ln(r/\sqrt{t})$ with a linear form, $f(x) = mx + c$, gives $\alpha\simeq 0.95$ \blue{(red dashed)}. We also show the curve with slope one (blue) for comparison. (d) Scaled plot of the structure factor in log-log scale confirming the non-porod law: $S(k)\mathcal{L}(t)^{-d}$ as a function of  $k\mathcal{L}(t)$ where $d = 1$ also gives $\alpha\simeq0.95$. We show the corresponding Porod law behavior by the cyan dotted line. We have used $L=2048$ and $C=1.5$ and the quench $T$ to $T_c$. (e-h) Similar plots as (a-d) for $C = 4.0$. Each point represents an average of $10^3$ histories.}
	\label{coarseningCorrfunc}
\end{figure*}

\subsection{Coarsening properties}
To study coarsening, we start with a random initial condition, quench to various $T$, and study the static and dynamic properties of the system.

\subsubsection{The spatial correlation function and structure factor}
\label{spatialcorr}

One of the classic signatures of FDPO in a coarsening system is the presence of a cusp-singularity in the scaled spatial correlation function, $G(r,t)$, where $r$ is the spatial separation and $t$ is the time from the initial state, 
\begin{equation}
\label{corrdef}
G(r,t) =  \frac{1}{L} \left\langle \sum_{i=1}^L  S_i(t)S_{i+r}(t)\right\rangle,
\end{equation}
where $\langle \ldots\rangle$ denotes an average over initial conditions and histories. Figures \ref{coarseningCorrfunc}(a) and (e) show $G(r,t)$ for two different values of $C$. Theories of domain growth have established that there is a growing coarsening length, $\mathcal{L}(t)$ within which the system is effectively in equilibrium, and the correlation function follows scaling: $G(r,t) = f(r/\mathcal{L}(t))$ \cite{Puri2009,Bray1994}. This implies a data collapse for $G(r,t)$ as a function of $r/t^{1/z}$ for a suitably chosen $z$, where $\mathcal{L}(t) \sim  t^{1/z}$; here $z$ is the dynamic scaling exponent. Using the simulation data of $G(r,t)$, we obtain a data collapse to a master curve for $z=2$ (Figs. \ref{coarseningCorrfunc} b and f). Thus, for the TIDSI model, we obtain $\mathcal{L}(t) \sim  t^{1/z} = \sqrt{t}$. Note that the data collapse to the master curve does not occur at smaller $t$, which can be understood as follows. As we discuss in Sec. \ref{finitesizealpha}, the cusp exponent $\alpha$ is affected strongly by finite size effects in the steady state. These are reflected while coarsening by a lack of collapse if $\mathcal{L}(t)$ is not large enough.

As the length scale $\mathcal{L}(t)$ increases, it ultimately reaches the system size $L$, whereupon the system reaches equilibrium. To test for FDPO, we examine the slope, $\alpha$, of the equilibrium correlation function, $G(r|L)$, as a function of $r/L$, near the origin, $r/L\to0$. For the TIDSI-CL model, an analytical calculation \cite{Barma2019} of $G(r|L)$ along the critical line yields
\begin{equation}\label{analyticalcofr}
G(r|L) \approx 1 - \frac{1}{c - 1}\big(\frac{r}{L}\big)^{2-c},
\end{equation} 
and we read off the cusp exponent as $\alpha=2-c$.

As discussed in Sec. \ref{critline}, the value of $c$ has an upper bound of 1.172 in the TIDSI model. Since this bound is close to 1 and the predicted value is $\alpha=2-c$, we expect $\alpha$ will be close to 1 \cite{Barma2019,barma2023book}. To test for FDPO in the coarsening regime of the TIDSI model, we replace $L$ by $\mathcal{L}(t)$. As we have seen, $G(r,t)$ obtained from Monte Carlo simulations displays a scaling collapse. On plotting $[1-G(r,t)]$ as a function of $r/\sqrt{t}$ in a log-log plot, we observe a straight line for small $r/\sqrt{t}$, implying a power law variation; the slope of this line gives $\alpha$. Figures \ref{coarseningCorrfunc}(c) and (g) show these plots for the two values of $C$. The fits with a power law give $\alpha\simeq 0.96$ for $C=1.5$ and $\alpha=0.89$ for $C=4.0$. These values are close to those obtained from analytic theory for the TIDSI-CL model (Table \ref{alphatable}).

\begin{center}
	\begin{table}
			\caption{Comparison of the analytical values of $\alpha$ with that obtained in our simulations during coarsening when we quench the system from a high $T$ to $T_c$.}
		\label{alphatable}
		\begin{tabular}{|wc {0.8cm} |wc{2.0cm} |wc{0.8cm} | wc{2.1cm} |wc{2.1cm}|} 
			\hline\hline
			$C$ & Quench $T$ & $c$& $\alpha$ (Simulation) & $\alpha$ (Analytic) \\
			\hline
			1.5 & $T_c = 1.4304$ & 1.05 & 0.96 & 0.9514 \\ 
			\hline
			4.0 & $T_c = 3.6145$ & 1.11 & 0.89 & 0.8933 \\
			\hline\hline
		\end{tabular}
	\end{table}
\end{center}

We can estimate the value of $\alpha$ via another analysis: taking the Fourier transform of $G(r,t)$, we obtain the structure factor, $S({\bf k}, t)$, at wavevector $\mathbf{k}$,
\begin{equation}
S({\bf k}, t) = \frac{1}{L} \Big\langle \Big|\sum_{i = 1}^{L} S_i(t) e^{i{\bf k}\cdot \mathbf{r}_i} \Big |^2 \Big \rangle,
\end{equation}
where ${\bf r}_i$ is the position of the $i$th spin and $k = \frac{2\pi j}{L}$ with $j = 0,1,2 \ldots L/2$ \cite{gawlinski1985}. 
We scale the wave-vector by the coarsening length, $\mathcal{L}(t)$, and obtain $S(k,t) = \mathcal{L}(t)^d f(k\mathcal{L}(t))$, where $d=1$ is the dimension. The large $k$ behavior of the function $f$ has the form $f(k)\sim k^{-(d+\alpha)}$, where $\alpha = 1$ corresponds to the Porod law \cite{G.Porod1982,Shrivastav2014}. A log-log plot of $S(\mathbf{k},t)/\mathcal{L}(t)$ as a function of $k\mathcal{L}(t)$ yields a straight line of slope $2-\alpha$ when $k\mathcal{L}(t)\gg1$. We show these plots for two values of $C$ in Figs. (d) and (h). This gives us an alternative means of measuring the values of $\alpha$, which agree with the estimates in Table \ref{alphatable} and the analytic results.

\begin{figure}
	\centering
	\includegraphics[width=8.2cm]{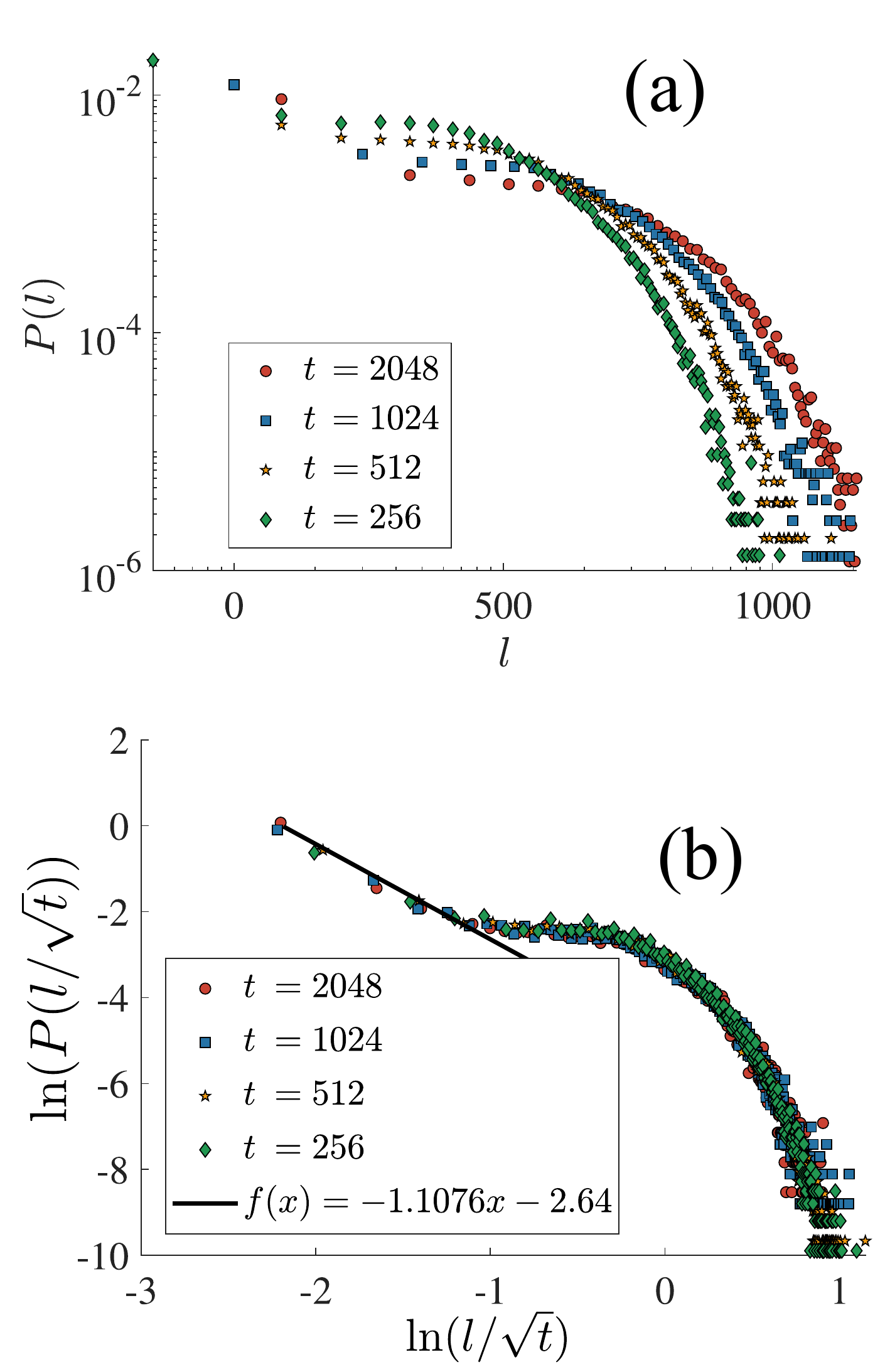}
	\caption{Domain size distribution, $P(l)$, during coarsening for $L = 2048$, $C = 4.0$, and a quench to $T_c=3.6145$. (a) $P(l)$ for various $t$. (b) $P(l)$ as a function of $l/\sqrt{t}$ collapses into a single curve. We fit the linear part with a straight line whose slope gives $c$. This value agrees with other estimates.}
	\label{dist_all_dom_coarsening}
\end{figure}

\subsubsection{Domain size distribution, $P(l)$, during coarsening}
Another typical characteristic of FDPO is a power law distribution of domain size, $l$ \cite{barma2023book,Chatterjee2006,Nagar2005,Das2000,Das2001}. Reference \cite{Barma2019} showed that for the TIDSI-CL model, the distribution, $P(l)$, in equilibrium follows $P(l)\sim l^{-c}$ for small $l$. For large $t$, we expect the system to equilibrate within the length scale of $\mathcal{L}(t)$, which would imply that the same law would hold for small $l$ even during coarsening. We have tested this prediction numerically. Figure \ref{dist_all_dom_coarsening}(a) shows $P(l)$ at different $t$ for $C=4.0$ and quench to $T_c$. Since $\mathcal{L}(t)\sim t^{1/2}$, we expect a data collapse for different $t$-curves when we plot $P(l)$ as a function of $l/\sqrt{t}$ (Fig. \ref{dist_all_dom_coarsening}b).
The power law decay at small $l$ becomes linear in the log-log plot (Fig. \ref{dist_all_dom_coarsening}b). The corresponding slope is very close to $c\simeq1.107$, consistent with the results obtained via $G(r,t)$ and $S(k,t)$ above.

\begin{figure}
	\centering
	\includegraphics[width=8cm]{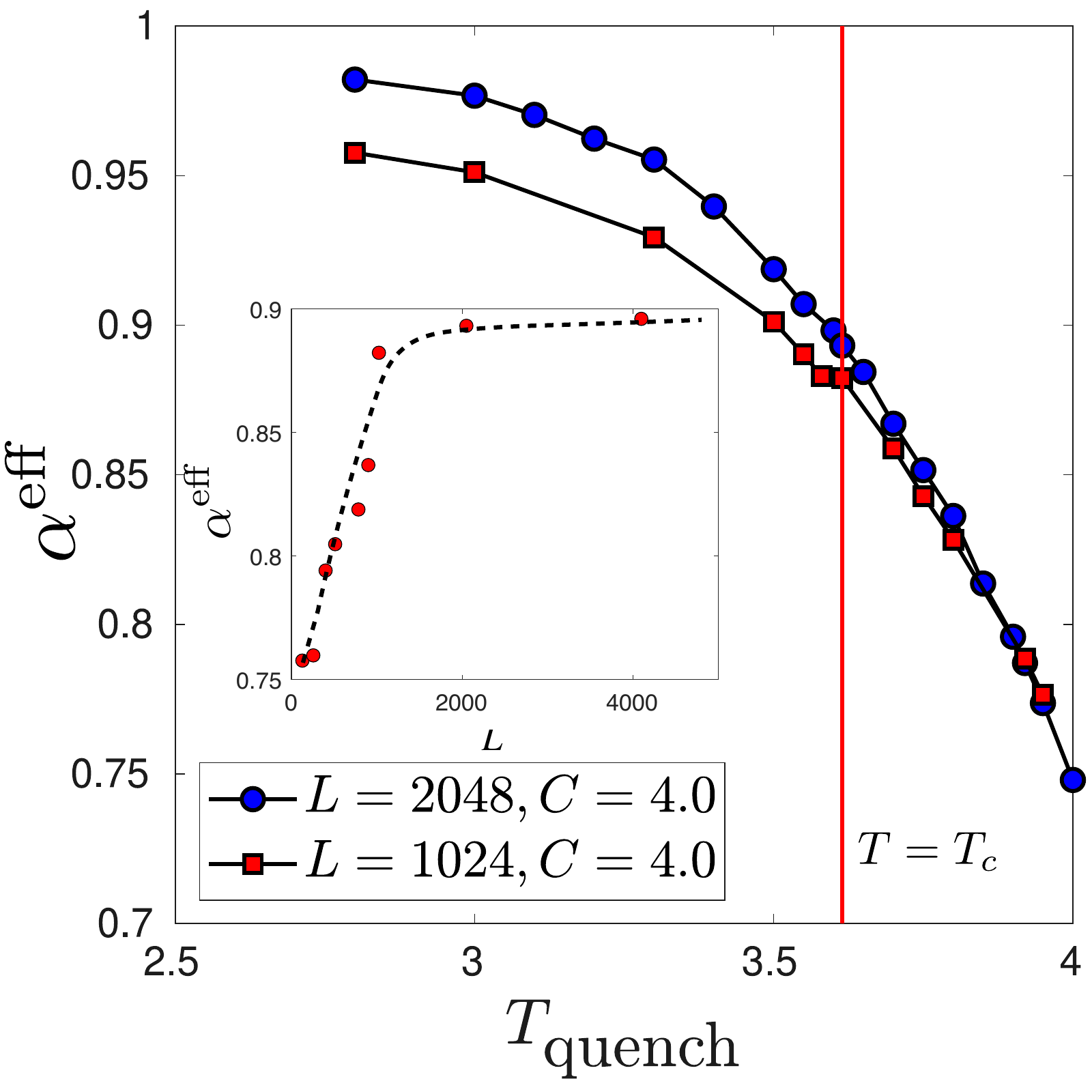}
	\caption{The effective cusp exponent, $\alpha^\text{eff}$, for various quench temperatures, $T_\text{quench}$. $\alpha^\text{eff}$ varies from analytical $\alpha$ when $T_\text{quench}$ differs from $T_c$. When $T_\text{quench}<T_c$, we obtain $\alpha^\text{eff}\simeq 1$; conversely, when $T_\text{quench}>T_c$, we obtained $\alpha^\text{eff}<\alpha$. The variation of $\alpha^\text{eff}$ is sharper for larger $L$. The value of $\alpha^\text{eff}$ becomes nearly independent of $L$ at large $L$. {\bf Inset}: Variation of $\alpha^\text{eff}$ with $L$ for $T_\text{quench}=T_c$. The dashed line is a guide to the eye showing that $\alpha^\text{eff}$ nearly saturates at higher $L$. Thus, we take the values of $\alpha^\text{eff}$ for $L=2048$ in the main figure as independent of system size.}
	\label{alphavsQuench}
\end{figure}

\begin{figure*}
	\centering
	\includegraphics[width=17.4cm]{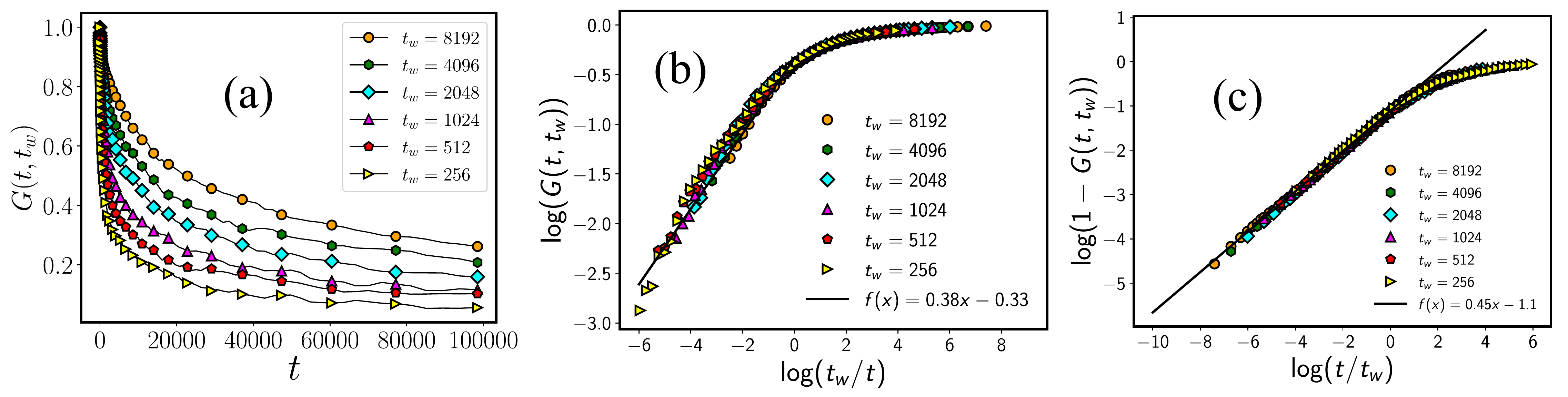}
	\caption{Aging autocorrelation function for the TIDSI model. (a) The plot of $G(t,t_w)$ as a function of $t$ for various $t_w$. (b) We plot $\ln G(t,t_w)$ as a function of $\ln (t_w/t)$ to verify the scaling prediction, Eq. (\ref{agingAutocorrgammaEq}). A linear fit with the early part of the data gives $\gamma\simeq 0.38$. (c) The plot of $\ln(1-G(t,t_w))$ as a function of $\ln(t/t_w)$ also shows data collapse to a master curve. Fit with a linear form with the early part of the data gives $\beta\simeq 0.45$ (Eq. \ref{agingAutocorrbetaEq}).}
	\label{betaextract}
\end{figure*}

\subsubsection{Finite size effect on the cusp exponent}
\label{finitesizealpha}
The correlation length $\xi$ is a function of $c$ in the TIDSI-CL model and is predicted to diverge very strongly for $c$ close to 1. It can become much larger than $L$ as we approach the critical point, leading to strong finite-size effects on properties close to $T_c$. Bar {\it et al.} have shown that $\xi  \sim (T-T_c)^{-\nu}$, with $\nu$ given by \cite{Bar2016}
\begin{equation}
\nu= 
\begin{cases}
\frac{1}{c-1},& \text{if } 1 \leq c < 2\\
1,              & \text{if } c > 2.
\end{cases}
\end{equation}

As discussed in Sec. \ref{critline}, the physical range of the TIDSI model is $c \leq 1.172$. In our simulations, for $C=4.0$, we have $c =1.107$, which would correspond to $\nu \sim 10$, implying that $\xi$ rapidly becomes larger than $L$, in a broad range around the critical point. As a result, even when the quench temperature satisfies $T_\text{quench}> T_c$, we see FDPO-like behavior in a large range of $T_\text{quench}$. However, we observe a variation of the cusp exponent, which leads us to define an effective cusp exponent, $\alpha^\text{eff}$, which depends on $T_\text{quench}$. We find $\alpha^\text{eff}$ is a smooth function of $T_\text{quench}$ as shown in Fig. (\ref{alphavsQuench}). We see,
\begin{equation}
\alpha^\text{eff}
\begin{cases}
\simeq 1, &\text{ when } T_\text{quench}<T_c,\\
=\alpha, &\text{ when } T_\text{quench}=T_c,\\
<\alpha,  &\text{ when } T_\text{quench}>T_c.
\end{cases}
\end{equation}

In the $L\to\infty$ limit, we expect $\alpha=1$ in the ordered phase and $\alpha=0$ in the disordered phase. Presumably, the smooth variation of $\alpha^\text{eff}$ in Fig. \ref{alphavsQuench} is an effect of finite size. The variation in $\alpha^\text{eff}$ becomes sharper with higher $L$. 
\r{Furthermore, for $T_\text{quench}=T_c$, the inset of Fig. \ref{alphavsQuench} shows that $\alpha^\text{eff}$ varies rapidly at small $L$ and weakly at larger $L$ approaching the analytic $\alpha$ as $L$ becomes large. Similar system-size dependence at small $L$ also appears in the coarsening data presented in Sec. \ref{spatialcorr}. At small $t$, the system equilibrates within a small length scale $\mathcal{L}(t)$. However, since $\alpha^\text{eff}$ is different from the analytical value at small $\mathcal{L}(t)$ due to this finite size effect, we see the data collapse to the master curve only at larger $t$ (Fig. \ref{coarseningCorrfunc}).
In addition, the inset of Fig. \ref{alphavsQuench} shows that $\alpha^\text{eff}$ nearly saturates beyond $L\approx 2000$. Therefore, we ignore the system size dependence in the values of $\alpha^\text{eff}$ for $L=2048$ in Fig. \ref{alphavsQuench} and assume the variation is an effect of $T_\text{quench}$ alone.}

When $T_\text{quench}=T_c$, the system would reach steady state when $\mathcal{L}(t)\simeq L$, and the $t$-dependence would vanish. However, in our coarsening studies, we have confined $t$ to values such that $\mathcal{L}(t)$ is not more than 10\% of $L$. For higher values of $T_\text{quench}$, $\xi$ is smaller than $L$, and we expect the system to equilibrate much earlier. We have verified that for such values of $T_\text{quench}$, say $T_\text{quench}=4$ when $C=4$ (Fig. \ref{alphavsQuench}), $G(r,t)$ as a function of $r/\sqrt{t}$ shows data collapse up to specific values of $t$, and then deviates from it. The breakdown of the coarsening scaling with $t$ is consistent with the system reaching an equilibrium disordered state. We will show below that $\alpha^\text{eff}$ also plays a critical role in understanding the equilibrium properties of the TIDSI model.

\subsubsection{Aging auto-correlation function }
\label{agingautocorr}
We also studied dynamical properties via the two-point aging auto-correlation function, defined as
\begin{equation}
G(t,t_w) =  \frac{1}{L} \left\langle \sum_{i=1}^L S_i(t_w) S_i(t + t_w) \right\rangle,
\label{agingAutocorrEq}
\end{equation}
where $t$ is the time difference and $t_w$ is the waiting time after the quench from the high-temperature phase. There is no time-translational invariance during aging, and $G(t, t_w)$ depends on both $t$ and $t_w$. We know from the coarsening studies of systems around ordinary critical points that $G(t,t_w)$ is a function of $t_w/t$ when $1 << (t_w, t) << L^z$ \cite{Bray1994}. When $t >> t_w$, this function has a power law decay,
\begin{equation}
G(t,t_w) \sim \Big (\frac{t_w}{t}\Big)^{\gamma}.
\label{agingAutocorrgammaEq}
\end{equation}
In the other limit, when $t_w >> t$, the function has a cusp like form given by,
\begin{equation}
G(t,t_w) \sim m^2\Big [1 - b_1 \bigg(\frac{t}{t_w}\bigg)^{\beta}\Big],
\label{agingAutocorrbetaEq}
\end{equation}
where $\gamma$ and $\beta$ are two exponents,  $m^2 = 1$ for our system, and $b_1$ is a constant.

Similar scaling forms for the autocorrelation function have been found for other systems exhibiting FDPO, and analytical and numerical estimates of the exponents have been obtained \cite{Chatterjee2006}. We investigated the behavior of $G(t,t_w)$ for three values of $C$ and have shown representative plots for $G(t,t_w)$ for $C=4$ in Fig. \ref{betaextract}(a). We show the scaling of $G(t,t_w)$ as a function of $t_w/t$ in Fig. \ref{betaextract}(b) and of $t/t_w$ in Fig. \ref{betaextract}(c); the lines represent the fits with Eqs. (\ref{agingAutocorrgammaEq}) and (\ref{agingAutocorrbetaEq}) to extract $\gamma$ and $\beta$ respectively. The plots for other values of $C$ are similar. 

\begin{center}
	\begin{table}
		\caption{Values of the scaling exponents: $\alpha$, $\gamma$, and $\beta$}
		\begin{tabular}{|wc{1cm}| wc{2.3cm}| wc {2.4cm} | wc{1cm}| wc{1cm}| }
			\hline \hline
			$C$ & $\alpha$ (Analytic) & $\alpha$ (Simulation) & $\gamma$& $\beta$ \\
			\hline
			1.5 & 0.9514 & 0.96 & 0.32& 0.49 \\ 
			\hline
			3.0 & 0.9089 & 0.92 & 0.34& 0.47\\ 
			\hline
			4.0 & 0.8933 & 0.89 & 0.38& 0.45 \\
			\hline\hline
		\end{tabular}
       \label{tab:betaval}
	\end{table}
\end{center}  

Table \ref{tab:betaval} shows the values of $\gamma$ and $\beta$, along with the values of $\alpha$. Note the opposite trends of $\gamma$ and $\beta$: with increasing $C$, $\gamma$ increases whereas $\beta$ decreases.
Furthermore, Table \ref{tab:betaval} suggests that the exponent $\beta \simeq \frac{\alpha}{2}$. We can relate this to the steady state properties of $G(r,L)$ \cite{Barma2019},
\begin{equation}\label{steadystateCofr}
	G(r, L) \sim m^2 \left[1 - b \left(\frac{r}{L}\right)^\alpha\right],
\end{equation}
where we have used the relation $\mathcal{L}(t_w)\sim t_w^{1/2}$ and replaced $L$ by $\mathcal{L}(t_w)$ and $r$ by $\sqrt{t}$ as $z=2$ for our system. Therefore, from Eq. (\ref{steadystateCofr}), we obtain Eq. (\ref{agingAutocorrbetaEq}) with $b$ being a constant proportional to $b_1$ and $\beta=\alpha/2$.

 \begin{figure}
	\centering
	\includegraphics[width=8cm]{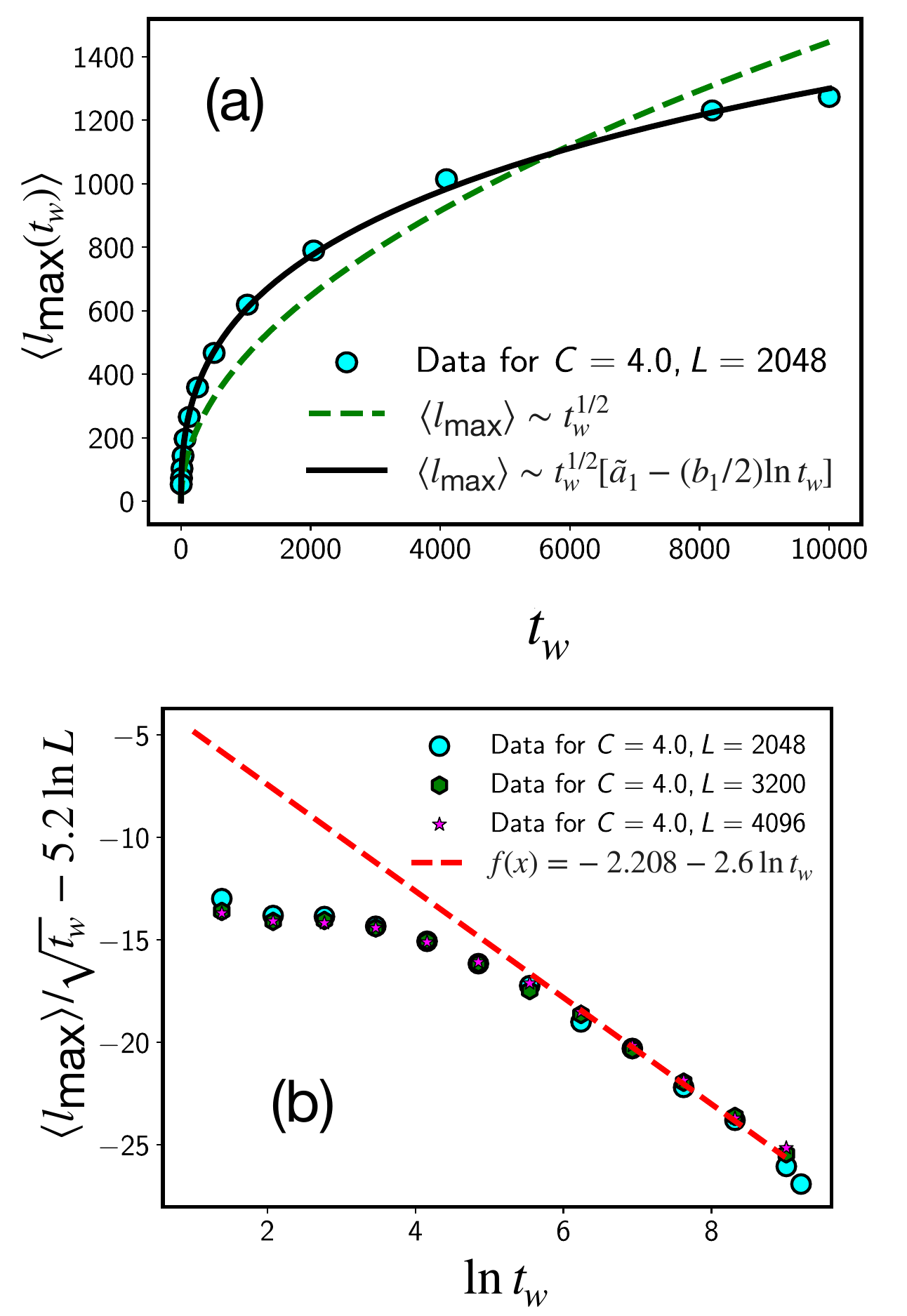}
	\caption{Corrections to $\langle l_\text{max}(t_w) \rangle$ during coarsening. (a) $\langle l_\text{max}(t_w) \rangle$ as a function of $t_w$, points represent simulations of TIDSI model with $C= 4.0, L = 2048$ and $T_\text{quench} = T_c (= 3.6145)$ and fixed $L$. The solid line is the fit with $\langle l_\text{max}\rangle =t_w^{1/2} [\tilde{a}_1 - (b_1/2) \ln t_w]$, where $\tilde{a}_1=a_1+b_1\ln L$ with $a_1=-2.208$ and $b_1=5.2$ are constants. We also show the fit with $\langle l_\text{max}\rangle\sim t_w^{1/2}$ (dashed line) for comparison.
		(b) To test the corrections for varying $L$, we plot $\frac{\langle l_\text{max}(t_w) \rangle}{\sqrt{t_w}} - 5.2 \ln L$ as a function of $\ln t_w$ for different $L$. The data collapse to a master curve, and the agreement with Eq. (\ref{FSC}) at large $t_w$, confirms the importance of the logarithmic corrections in Eq. (\ref{FSC}).}
	\label{correction_lmax_coarsening}
\end{figure}

\subsubsection{Finite size correction for the average size of the largest cluster}
\label{averageclusteraging}

A crucial aspect of FDPO is that the size of the largest cluster, $l_\text{max}$, scales with the system size $L$ \cite{barma2023book,Das2000,Chatterjee2006}. Thus, $l_\text{max}$ has special significance for the time-evolution. A recent work, Ref. \cite{Das2023}, has presented an analytical argument for the finite size correction to the mean largest cluster size of the coarse-grained depth (CD) model during coarsening as,
{\begin{equation}\label{FSC}
\langle l_\text{max}(t_w) \rangle =  t_w^{1/z} \Big [a_1 + b_1 \ln(L) - \frac{b_1}{z} \ln(t_w)\Big],
\end{equation}
where $a_1$, and $b_1$ are constants, and $z$ is the scaling exponent (in our case $z =2$). We have separately tested for the two types of corrections in Eq. (\ref{FSC}) due to finite $L$ and finite values of $t_w$. We first verified the logarithmic correction of $t_w$. For a fixed system size, $L = 2048$, we compute $l_\text{max}$ as a function of $t_w$. Figure \ref{correction_lmax_coarsening}(a) shows the comparison of the simulation data (symbols)  with (solid line) and without (dashed line) the logarithmic corrections. It is evident that it is essential to include the correction in Eq. (\ref{FSC}) for the $\langle l_\text{\text{max}}\rangle$ data. We next verified the finite $L$ correction by computing $\langle l_{max}(t_w) \rangle$ for three different $L$. Figure \ref{correction_lmax_coarsening}(b) shows the plot of $\frac{\langle l_{\text{max}}(t_w) \rangle}{\sqrt{t_w}} - 5.2 \ln L$ as a function of $\ln t_w$. The data collapse for this specific form, where we have subtracted $\ln L$, shows that the logarithmic correction is important, and the simulation data are consistent with the finite size scaling correction, Eq. (\ref{FSC}) \cite{Das2023}.

\begin{figure*}
	\centering
	\includegraphics[width=15cm]{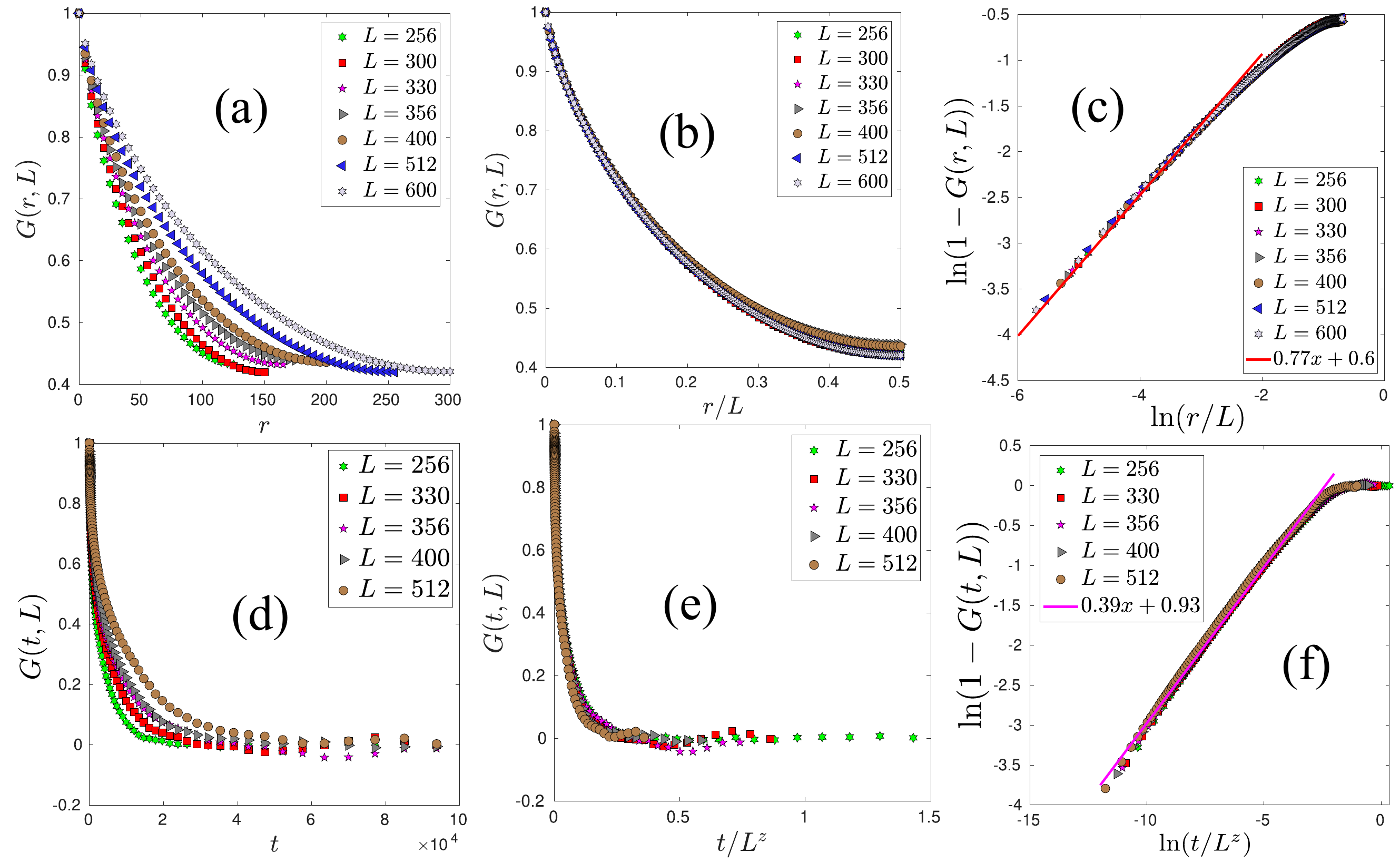}
	\caption{The steady-state correlation functions of the TIDSI model. (a) The spin-spin spatial correlation function, Eq. (\ref{SS_spa_corr}), for varying system sizes with $C = 4$ and $T = 3.95$. (b) Plot of $G(r,L)$ as a function of $r/L$ shows data collapse to a master curve. (c)The plot of $\ln(1-G(r,L))$ as a function of $\ln(r/L)$ for the same data as in (a) shows data collapse to a master curve. We obtain $\alpha^\text{eff}$ from the initial part of the data by fitting a straight line whose slope gives $\alpha^\text{eff}$. The value of $\alpha^\text{eff}$ agrees that in Fig. \ref{alphavsQuench}. (d) The autocorrelation function, $G(t,L)$, as a function of $t$ for five different system sizes at $T=3.95$ and $C=4$. (e) Plot pf $G(t,L)$ as a function of $t/L^z$ shows data collapse. (f) The plot of $\ln(1-G(t,L))$ as a function of $\ln(t/L^z)$ with $z=2$ shows data collapse to a master curve. The slope for the initial part of the data gives $\beta$ that agrees with the expected value of $\beta=\alpha^\text{eff}/2$. Each point on this plot represents averages over at least 24 ensembles and 100 well-separated initial times.
	}
	\label{SteadyStateProp1}
\end{figure*}

\subsection{Steady state statistics}
We now present our simulation results for the steady-state properties. Note that the presence of the long-range term in $\mathcal{H}$, Eq. (\ref{Hamiltonian}), increases the computation time significantly and makes equilibration of the TIDSI model for large system sizes time-consuming. To alleviate this difficulty, we simulated relatively smaller systems and used the broadness of the critical regime (Sec. \ref{finitesizealpha}) to justify simulating the system slightly above $T_c$. We equilibrated the system for at least $10^6$ MC times before collecting data.

\subsubsection{Spin-Spin spatial correlation function}
We first study the behavior of the spin-spin spatial correlation function, $G(r,L)$,
\begin{equation}
\label{SS_spa_corr}
G(r,L) =\overline{\frac{1}{L}\left\langle\sum_{i=1}^L \langle S_i(t)S_{i+r}(t)\right\rangle},
\end{equation}
The $\langle \ldots \rangle$ denotes the average over ensembles and the overbar denotes that over $t$. For finite-size systems, $G(r,L)$ shows scaling as a function of $r/L$ in the limit of $r\to\infty$ and $L\to\infty$ with finite $r/L$.
For systems showing FDPO in the steady state, in the limit of $\frac{r}{L} << 1$, $G(r,L)$ decays as a power law with a cusp exponent $\alpha^\text{eff}$ and an intercept $m^2$:
\begin{equation}
G(r,L) = m^2\Big [1 - a \bigg(\frac{r}{L}\bigg)^{\alpha^\text{eff}}\Big],\,\,\, \text{for } \frac{r}{L} \ll 1,
\label{SS_spacorrEq}
\end{equation}
where $a$ is a constant and $m^2$ is a measure of long-range order in the system; $m^2=1$ for our system. Figure \ref{SteadyStateProp1}(a) shows $G(r,L)$ as a function of $r$ for various $L$ for fixed $T=3.95$ and $C=4$. When we plot $G(r,L)$ as a function of $r/L$, we observe data collapse to a master curve (Fig. \ref{SteadyStateProp1}b). As for coarsening, discussed in Sec. \ref{spatialcorr}, we calculate $\alpha^\text{eff}$ (see Sec. \ref{finitesizealpha}) from the plot of $\ln(1-G(r,L))$ vs. $\ln({r}/{L})$ by fitting a straight line to the early part of the curve (Fig. \ref{SteadyStateProp1}c). We find that $\alpha^\text{eff}=0.77$; this is consistent with the critical-like behavior with an effective cusp exponent for temperatures slightly above $T_c$ and agrees with the earlier value shown in Fig. \ref{alphavsQuench}.

\subsubsection{Time auto-correlation function}
We study the dynamical properties in the steady state via the two-time auto-correlation function, $G(t,L)$, 
\begin{equation}
G(t,L) = \overline{\frac{1}{L} \left\langle \sum_{i=1}^L S_i(t_0) S_i(t_0+t) \right\rangle}
\label{SSAutocorrEq}
\end{equation}
where $t_0$ is the time origin. Figure \ref{SteadyStateProp1}(d) shows $G(t,L)$ as a function of $t$ for different values of $L$ for the parameters $T = 3.95$ and $C = 4$. In the steady state, $G(t, L)$ shows scaling as a function of $\frac{t}{L^z}$, where $z=2$ is the dynamic exponent. As for other systems displaying FDPO \cite{Chatterjee2006}, this function exhibits scaling, with a cusp in the small argument limit (Fig. \ref{SteadyStateProp1}e). Following a similar argument as in Sec. \ref{agingautocorr}, we can write down the scaling form of the autocorrelation function using Eq. (\ref{SS_spacorrEq}) as
\begin{equation}
G(t,L) = m^2\Big [1 - b \bigg(\frac{t}{L^z}\bigg)^{\beta}\Big], \,\,\, \text{for} \,\,\frac{t}{L^{z}} \rightarrow 0,
\label{SS_AutocorrEq}
\end{equation}
where $b$ is a constant. We find the cusp exponent $\beta$ from the plot of $\ln(1-G(t,L))$ as a function of  $\ln(t/L^z)$ by fitting a straight line with the initial part of the curve (Fig. \ref{SteadyStateProp1}f). The value of $\beta\simeq0.39$ is consistent with the argument in Sec. \ref{agingautocorr} with $\beta=\alpha^\text{eff}/2$, and also consistent with the predictions of Ref. \cite{Barma2019}.

\begin{figure}
	\centering
	\includegraphics[width=8cm]{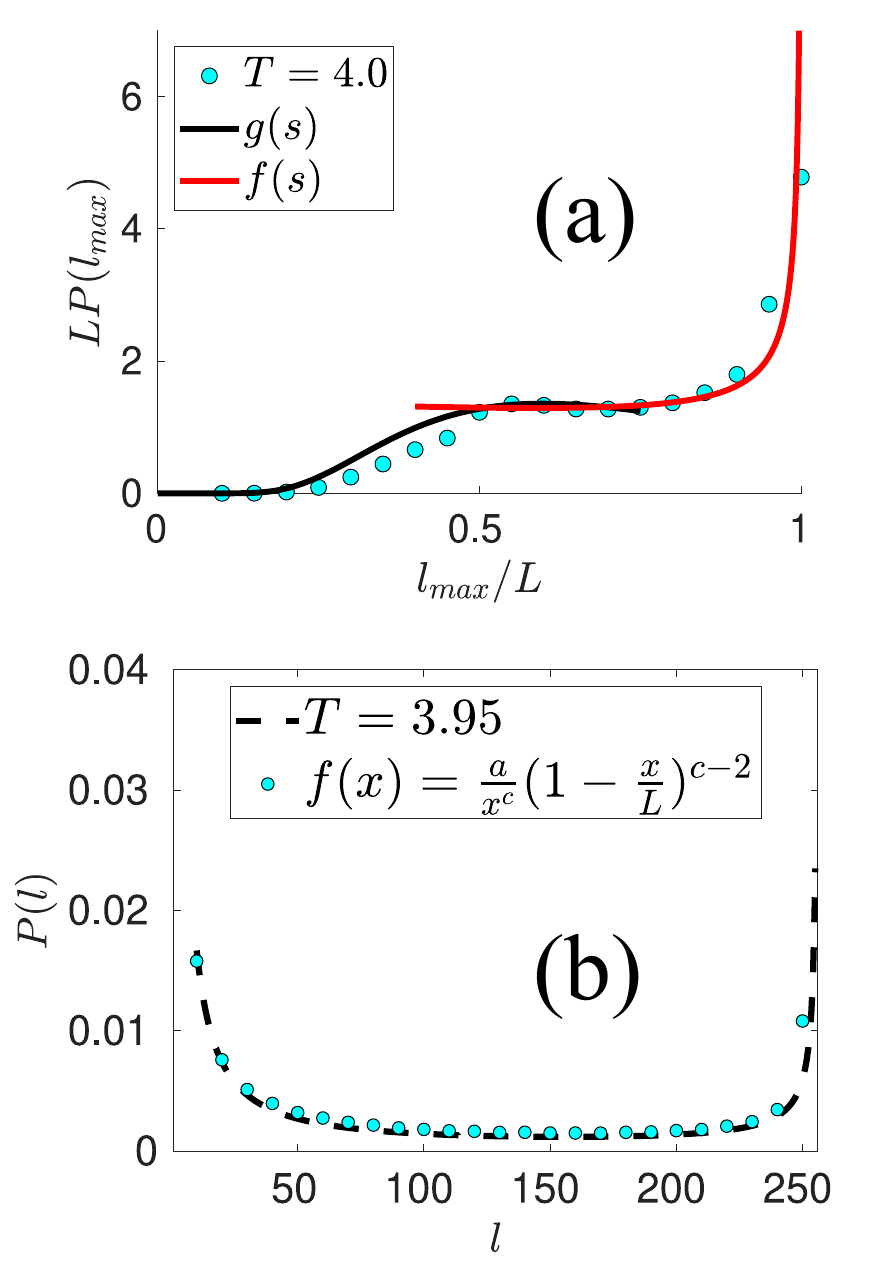}
	\caption{Steady-state distribution of domain sizes. (a) The scaled PDF of $l_\text{max}$. Points represent simulation data of the TIDSI model with $C = 4.0, T = 4.0$, and $L = 400$. The line represents the plot of $g(s)$ (Eq. \ref{largestdomdistEq}) with $c = 1.252$. (b) The distribution of domain sizes, $P(l)$, in steady state. Points represent simulation data of the TIDSI model with $C= 4.0$, $T = 3.95$, $L = 256$, and the line shows Eq. (\ref{alldomdistEq}) with $c = 1.22$ and $a = 0.27$.}
	\label{SteadyStateProp}
\end{figure}

\subsubsection{Domain size distribution}
Clusters in FDPO are very dynamic and fluctuate strongly. Bar {\it et al.} studied analytically the extreme-value statistics of the TIDSI-CL model and obtained the distribution of $l_\text{max}$ \cite{Bar2016}. In the FDPO regime, the distribution $g$ follows
\begin{equation}
g(s) \approx \begin{cases}
\gamma_0e^{-\alpha_0/s}s^{c-4}(1-s\frac{2-c}{\alpha_0}) + \mathcal{O}(e^{-\alpha_1/s}), & s \rightarrow 0 \\
\gamma_1(1-s)^{2c-3} + (1-s)^{2c-2}, & s \rightarrow 1
\end{cases}
\label{largestdomdistEq}
\end{equation}
where, $s = l_\text{max}/L$, $\gamma_0 = \frac{\pi \alpha_0^2 e^{-\alpha_0}}{(c-1)\sin[\pi(c-1)]}$, $\gamma_1 = -2(c-1)\frac{\Gamma(c-1)}{\Gamma(1-c)\Gamma(2c-1)}$, and $-\alpha_0$ is the single negative real zero of $\Gamma(1-c,w) - \Gamma(1-c)$ as a function of $w$. We have tested this analytic result in our simulation. Figure \ref{SteadyStateProp}(a) shows the simulation data (symbols), while the lines are plots of Eq. (\ref{largestdomdistEq}) with $c=1.25$. This value of $c$ is consistent with the coarsening results, Fig. \ref{alphavsQuench}, for $\alpha^\text{eff}$ (note that the bound of $c^*$ need not apply for $T>T_c$). The distribution shows a singularity at $l_\text{max}/L = 0.5$, where the two forms cross over (Fig. \ref{SteadyStateProp}a).
We note that multiplying the early part of $g(s)$ in Eq. (\ref{largestdomdistEq}) with a factor of $3/4$ would give a better agreement with the simulation data.

We have also tested the prediction of the domain size distribution, $P(l)$, in a finite size system \cite{Barma2019}:
\begin{equation}
P(l) \approx \frac{y_c}{l^c}\Big(1 - \frac{l}{L}\Big)^{c-2}
\label{alldomdistEq}
\end{equation}
where $y_c = \frac{1}{\zeta(c)}$. We show $P(l)$ for a system with $L = 256$, $T = 3.95$, and $C = 4$ in Fig. \ref{SteadyStateProp}(b). We obtain $c=1.22$ by fitting Eq. (\ref{alldomdistEq}) with the data; this value is very close to $(2-\alpha^\text{eff})$, discussed in Sec. \ref{finitesizealpha}, and shown in Fig. \ref{alphavsQuench}. Thus, the simulation data bear out the analytic prediction.

\section{Conclusions and Discussion}

The TIDSI model is the first example of an equilibrium model which shows FDPO along the phase boundary. FDPO is characterized by a cusp of the scaled correlation function and extensive fluctuations of the order parameter. Unlike other known instances of FDPO (in nonequilibrium systems) there is a continuous variation of the cusp exponent along the phase boundary. The transition across the phase boundary in the TIDSI model is a mixed order transition (MOT), with a discontinuity of the order parameter and a diverging correlation length \cite{Bar2014,Bar2014PRL,Bar2016}. Parenthetically, we remark that in nonequilibrium systems exhibiting FDPO, there is a jump of the order parameter across the transition, but a divergent correlation length has not been demonstrated, so the likely MOT character in such cases remains to be firmly established.

The analytical calculations of Ref. \cite{Barma2019} were performed on the TIDSI-CL model that represents the Hamiltonian in terms of cluster lengths. In this paper, we performed a Monte Carlo study of the original TIDSI model, aimed at clarifying the connection with the results for the TIDSI-CL model. While the TIDSI-CL model is defined for all $c>1$, we demonstrated that the physically accessible range for the TIDSI model is narrowed down to $1\leq c\leq c^*$, where $c^*\simeq 1.172$. Within this range, our simulations of the TIDSI model provide the first numerical verification of the analytical results for the TIDSI-CL model. Note that since $c^*<2$, the TIDSI model always displays FDPO for ferromagnetic nearest-neighbor interactions. 

Since the value of $c$ is restricted to be close to 1, it follows that the correlation length exponent $\nu=1/(c-1)$ is large, implying a rapid growth of the correlation length, $\xi$. This leads to a broad range of temperature in which critical-like behavior is observed, even when $T$ is higher than $T_c$.

An intriguing result of our simulations is the finite-size effect on the cusp exponent as a function of the distance of the quench temperature, $T_\text{quench}$, from the critical point, $T_c$. For an infinite system, we expect $\alpha=1$ for $T_\text{quench}<T_c$ and $\alpha=0$ for $T_\text{quench}>T_c$. Instead, we find a smooth variation of $\alpha$ as a function of $T_\text{quench}$ that leads us to define an effective value $\alpha^\text{eff}$ (Fig. \ref{alphavsQuench}). The variation becomes sharper as the system size $L$ increases. The system-size effect on $\alpha^\text{eff}$ becomes weaker at larger $L$. 
We also studied the coarsening dynamics of the TIDSI model after a sudden quench from a high temperature. During coarsening, the system is in equilibrium within a length $\mathcal{L}(t)\sim t^{1/2}$ and the properties are consistent with those in equilibrium when we replace $L$ by $\mathcal{L}(t)$. For example, the cusp exponent during coarsening agrees with the analytical result in equilibrium \cite{Barma2019}. 
Since $\alpha^\text{eff}$ varies significantly at smaller $\mathcal{L}(t)$, we find the data collapse only at larger $t$ when $\mathcal{L}(t)$ grows appreciably. Conversely, the significant variation of $\alpha^\text{eff}$ with $T_\text{quench}>T_c$ rationalizes our equilibrium results where we can only equilibrate the system for $T>T_c$, where it continues to show critical-like behavior as the correlation length is extremely large.

We have also studied the aging dynamics for a critical quench via the two-point autocorrelation function, $G(t,t_w)$. We showed that $G(t,t_w)$ exhibits scaling for two distinct regimes characterized by two exponents, $\gamma$ when $t\gg t_w$ and $\beta$ when $t\ll t_w$. We argued and numerically verified that $\beta=\alpha/2$. Since $\beta$ characterizes the decay of the auto-correlation function, we note that FDPO provides a mechanism for inducing stretched exponential relaxation \cite{vaibhav2020,pareek2023}.

It would be interesting to extend the model to higher dimensions by including long-range interactions within domains of like-spins and to investigate the specific conditions (such as the nature of the long-range interaction or the exponent of the power-law from) under which the FDPO behavior survives. Interestingly, several biologically relevant theoretical models, such as active matter comprised of self-propelled particles \cite{sriramreview,sriramrmp,activereview} or confluent models of epithelial monolayers \cite{graner1992,glazier1993,farhadifar2007,sadhukhan2021,sadhukhan2024}, possess implicit long-range interactions that may impact the phase ordering kinetics in these systems.

\section{Acknowledgements}
We thank David Mukamel, Arghya Das, Satya N Majumdar, Gregory Schehr, Varsha Banerjee and Sakuntala Chatterjee for discussions. We acknowledge the support of the Department of Atomic Energy, Government of India, under Project Identification No. RTI 4007. SKN thanks SERB for grant via SRG/2021/002014 and MB acknowledges the support of the Indian National Science Academy.

\appendix

\section{Equivalence of the results with different boundary conditions}
\label{comp}

\begin{figure}[t]
	\centering
	\includegraphics[width=8.5cm]{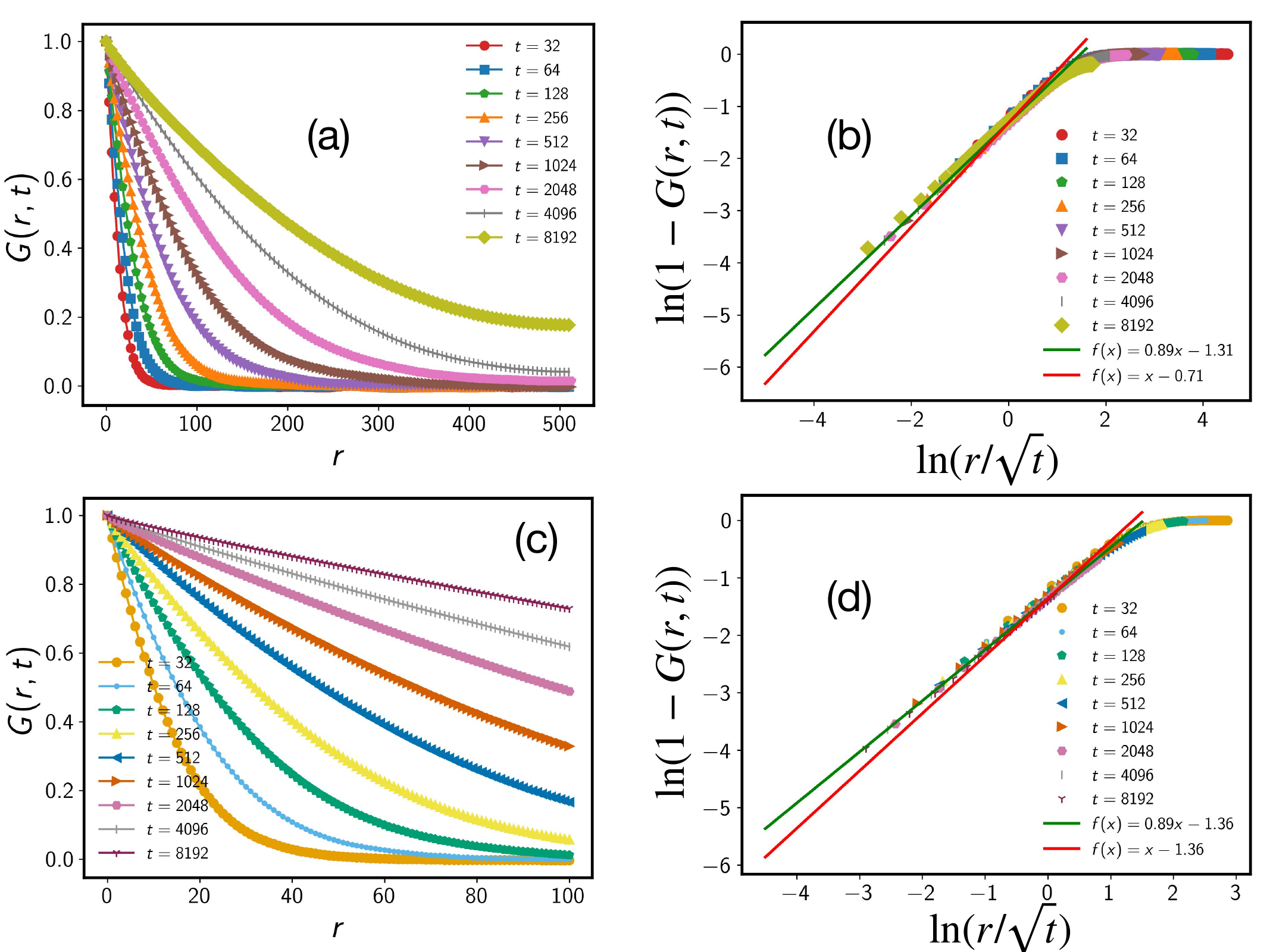}
	\caption{TIDSI model with different boundary conditions during coarsening at $T = T_c$. (a) $G(r,t)$ vs. $r$ for various $t$ for the system with PBC. As $t$ increases, the decay of the curves become slower. (b) Fitting of the small $r$ data of $\ln(1-G(r,t))$ as a function of $\ln(r/\sqrt{t})$ with a linear form, $f(x) = \alpha x + c$, gives $\alpha\simeq 0.89$. (c) and (d) are same as in (a) and (b) but for free boundary conditions (FBC). The linear fits in (b) and (d) gives the same value of $\alpha$, demonstrating that the specific boundary condition does not change the qualitative results. We have used $L=1024$ and $C=4$ for these simulations. }
	\label{diffboundary}
\end{figure}

As mentioned in Sec.~\ref{modeldes}, whereas the simulation results of the TIDSI model do depend on the boundary conditions, the value of the cusp exponent for correlation functions does not differ much when we use the periodic boundary condition (PBC) as opposed to a free boundary condition (FBC). To demonstrate this equivalence, we show the evolving spatial correlation functions, $G(r,t)$, after a sudden quench to $T = T_c$ for two systems with different boundary conditions. For the system with PBC (Figs. \ref{diffboundary} a and b), we obtained $G(r,t)$ up to a maximum separation of $r = L/2$ (Fig. \ref{diffboundary}a). On the other hand, for FBC (Figs. \ref{diffboundary}c and d), we find that boundary effects are minimal if we restrict spatial separation up to $r=100$ at the center for a system size $L=1024$. Figures~\ref{diffboundary}(a) and (c) show $G(r,t)$ as a function of $r$ for various $t$ as defined in Eq.(\ref{corrdef}) for the PBC and the FBC respectively. Figures \ref{diffboundary}(b) and (d) show the same data for $\ln(1-G(r,t))$ as a function of $\ln(r/\sqrt{t})$. A linear fit, $f(x) = mx + c$, with the data, gives $\alpha\simeq 0.89$, the same value for both the PBC and FBC, as shown in Fig.~\ref{diffboundary}(b) and (d), respectively, demonstrating that the results do not depend on the specific boundary condition.

\noindent
All data supporting the findings of this study are available in Ref.~\cite{Github}.

\bibliography{ref.bib}

\end{document}